\documentstyle[12pt]{article}

\begin{document}
\thispagestyle{empty}

\begin{center}
\LARGE \tt \bf{Turbulent ${\alpha}$-effect in twisted magnetic flux
tubes dynamos in Riemannian space}
\end{center}

\vspace{1.5cm}

\begin{center} {\large L.C. Garcia de Andrade \footnote{Departamento de
F\'{\i}sica Te\'{o}rica - Instituto de F\'{\i}sica - UERJ

Rua S\~{a}o Fco. Xavier 524, Rio de Janeiro, RJ

Maracan\~{a}, CEP:20550-003 , Brasil.E-mail:garcia@dft.if.uerj.br}}
\end{center}

\vspace{2.0cm}

\begin{abstract}
Analytical solution of first order torsion ${\alpha}$-effect in
twisted magnetic flux tubes representing a flux tube dynamo in
Riemannian space is presented. Toroidal and poloidal component of
the magnetic field decays as $r^{-1}$ , while grow exponentially in
time. The rate of speed of the helical dynamo depends upon the value
of Frenet curvature of the tube. The $\alpha$ factor possesses a
fundamental contribution from constant torsion tube approximation.
It is also assumed that the curvature of the magnetic axis of the
tube is constant. Though ${\alpha}$-effect dynamo equations are
rather more complex in Riemann flux tube coordinates, a simple
solution assuming force-free magnetic fields is shown to be
possible. Dynamo solutions are possible if the dynamo action is able
to change the signs of torsion and curvature of the dynamo flux tube
simultaneously.\vspace{0.5cm} \noindent {\bf
PACSnumbers:\hfill\parbox[t]{13.5cm}{02.40.Hw-Riemann geometries}}
\end{abstract}
\newpage
\section{Introduction}
 It is well-known that though, the astrophysical jets \cite{1} chaotic magnetic fields components decay as $r^{-1}$ and $r^{-2}$, unbounded
 magnetic fields \cite{1} may decay up to $r^{-3}$ at least. In this paper, the ${\alpha}$-effect in bounded system of a
 twisted magnetic flux tube, is shown to lead to a simple dynamo solution if one considers magnetic force-free fields \cite{1}.
 Despite the fact the dynamo equations are rather more complex than the cylindrical ones developed by Zeldovich, Ruzmaikin and
 Sokoloff \cite{2} rather simple solutions can be obtained on simple few cases. Despite of the success of the application
 of the numerical simulations to the dynamo problem \cite{3}
 in plasma astrophysics \cite{3} and in the stretch-twist-fold (STF)
 Vainshtein-Zeldovich \cite{2} mechanism, recently new dynamo
 analytical solutions have been found \cite{4} by using the
 conformal mapping technique in Riemannian manifolds from old Arnold
 cat dynamo metric \cite{5}. In this paper, the helical dynamo \cite{6} solution of the system of nonlinear system composed
 by the self-induction, solenoidal and force-free equations are
 found in the Riemann curved and torsioned flux tube coordinates.
 Dynamo solutions are obtained vorticity is stronger than torsion, and fast rate
 of the dynamo depends purely on the curvature of the thin twisted magnetic
 flux tubes in Riemannian, where MHD equations are linear in the magnetic field
 and nonlinear in the velocity flow. Recently, Hanasz and Lesch \cite{7} have also considered a conformal Riemannian
 metric in ${\cal{E}}^{3}$ to the galactic dynamo magnetic flux tubes.
 Pioneering work on the magnetic flux tubes as dynamos was done
 earlier by Schussler \cite{8}, however in his work tubes were untwisted and straight. The main advantage of the
 investigation of the isolated flux tube
 dynamo is that one is able to investigate the curvature and twist contributions of the tube to the dynamo action.
 Twist is actually related to the torsion of the magnetic axis of the tube, which makes the words strong torsion equivalent to
 strong twist, which physically is important to the twist-kink relation investigated by Alfven \cite{9}. Chaotic flows in magnetic
 fast dynamos \cite{10} mentioned above are however, kinematic dynamos in which velocity appears linearly in the dynamo
 equations, given a priori which makes their applications somewhat limited, this is one of the motivations which led us here
 to considered turbulent dynamo in magnetic flux tubes where non-linear velocity flows appear in the magnetic dynamo
 equations.  Helical dynamo here is understood as the one
 where the flow describes a circular helix where torsion and curvature are constants. The paper is organized as follows: In section
 II the ${\alpha}$-dynamo equations in the Riemann metric representing flux rope (twisted tubes) are obtained along with the In section III the
 approximate solution. In section III conclusions
 are given.
 \section{${\alpha}$-effect dynamo equations in Riemannian flux tubes}
In this section we shall be concerned with solving the MHD equations
in the curved coordinates of a thin twisted magnetic flux tube of
Riemann metric
\begin{equation}
ds^{2}=dr^{2}+r^{2}d{{\theta}_{R}}^{2}+{K^{2}}(s)ds^{2} \label{1}
\end{equation}
which represents a Riemann line element
\begin{equation}
ds^{2}=g_{ij}dx^{i}dx^{j} \label{2}
\end{equation}
if the tube coordinates are $(r,{\theta}_{R},s)$ \cite{1} where
${\theta}(s)={\theta}_{R}-\int{{\tau}ds}$ and $\tau$ is the Frenet
 torsion of the tube axis, $K(s)$ is given by
\begin{equation}
{K^{2}}(s)=[1-r{\kappa}(s)cos{\theta}(s)]^{2} \label{3}
\end{equation}
Since we are considered thin magnetic flux tubes, this expression
shall be taken as $K\approx{1}$ in future computations. In
curvilinear coordinates the Riemannian Laplacian operator
${\nabla}^{2}$ is
\begin{equation}
{\nabla}^{2}=\frac{1}{\sqrt{g}}{\partial}_{i}[\sqrt{g}g^{ij}{\partial}_{j}]
\label{4}
\end{equation}
where ${\partial}_{j}:=\frac{{\partial}}{{\partial}x^{j}}$ and
$g:=det{g_{ij}}$ where $g_{ij}$ are the covariant components of the
Riemann metric of flux rope.Let us now start by considering the MHD
field equations
\begin{equation}
{\nabla}.\vec{B}=0 \label{5}
\end{equation}
\begin{equation}
\frac{{\partial}}{{\partial}t}\vec{B}={\nabla}{\times}[{\alpha}\vec{B}]={\alpha}\lambda\vec{B}+{\nabla}{\alpha}{\times}{\vec{B}}
 \label{6}
\end{equation}
called the ${\alpha}$-dynamo equation \cite{2}. Where we have used
the force-free magnetic field equation
\begin{equation}
{\nabla}{\times}\vec{B}={\lambda}\vec{B}
 \label{7}
\end{equation}
in (\ref{6}), where ${\lambda}:=\frac{{\kappa}_{0}}{{\tau}_{0}}$,
defined in terms of the constant curvature and torsion, for future
computation convenience. To these equations one adds the sometimes
the ${\alpha}:=<\vec{v}.\vec{\omega}>$ parameter is constant but
here we shall be considering the more general case where it depends
on the radial and poloidal coordinate. Here
${\omega}:={\nabla}{\times}\vec{v}$ is the vorticity of the dynamo
flow. Equation (\ref{3}) represents the self-induction equation. The
vectors $\vec{t}$ and $\vec{n}$ along with binormal vector $\vec{b}$
form the Frenet holonomic frame, which obeys the Frenet-Serret
equations
\begin{equation}
\vec{t}'=\kappa\vec{n} \label{8}
\end{equation}
\begin{equation}
\vec{n}'=-\kappa\vec{t}+ {\tau}\vec{b} \label{9}
\end{equation}
\begin{equation}
\vec{b}'=-{\tau}\vec{n} \label{10}
\end{equation}
where the dash represents the ordinary derivation with respect to
coordinate s, and $\kappa(s,t)$ is the curvature of the curve, where
$\kappa=R^{-1}$. Here ${\tau}$ represents the Frenet torsion. The
gradient operator is
\begin{equation}
{\nabla}=\vec{t}\frac{\partial}{{\partial}s}+\vec{e_{\theta}}\frac{1}{r}\frac{\partial}{{\partial}{\theta}}+
\vec{e_{r}}\frac{\partial}{{\partial}r} \label{11}
\end{equation}
 Now we shall consider the analytical solution of the self-induction magnetic equation which represents a
 non-dynamo thin magnetic flux tube. Before the derivation of this result is obtained, we would like to point it out
 that it is not trivial, since the Zeldovich antidynamo theorem states that the two
 dimensional magnetic fields do not support dynamo action. Here, as is shown bellow, the flux tube axis
 possesses not only Frenet curvature, but torsion as well, and this
 last one vanishes in planar curves. The magnetic field does not
 possess a radial component and the magnetic field can be split inti its toroidal and poloidal components as
\begin{equation}
\vec{B}(r,s,t)={B_{\theta}}(t,r,{\theta}(s))+
B_{s}(r)\vec{t}\label{12}
\end{equation}
Now let us substitute the definition of the poloidal plus toroidal
magnetic fields into the self-induction equation,along with
expressions
\begin{equation}
\vec{e_{\theta}}=-\vec{n}sin{\theta}+\vec{b}cos{\theta} \label{13}
\end{equation}
and
\begin{equation}
\vec{e_{r}}=\vec{n}cos{\theta}+\vec{b}sin{\theta} \label{14}
\end{equation}
\begin{equation}
{\partial}_{t}\vec{e_{\theta}}={\omega}_{\theta}\vec{e}_{r}-{\partial}_{t}\vec{n}sin{\theta}+{\partial}_{t}\vec{b}cos{\theta}
\label{15}
\end{equation}
Considering the equations for the time derivative of the Frenet
frame given by the hydrodynamical absolute derivative
\begin{equation}
\dot{\vec{X}}={\partial}_{t}\vec{X}+[\vec{v}.{\nabla}]\vec{X}
\label{16}
\end{equation}
where $\vec{X}=(\vec{t},\vec{n},\vec{b})$ is used into the
expressions for the total derivative of each Frenet frame vectors
\begin{equation}
\dot{\vec{t}}={\partial}_{t}\vec{t}+[{\kappa}'\vec{b}-{\kappa}{\tau}\vec{n}]
\label{17}
\end{equation}
\begin{equation}
\dot{\vec{n}}={\kappa}\tau\vec{t} \label{18}
\end{equation}
\begin{equation}
\dot{\vec{b}}=-{\kappa}' \vec{t} \label{19}
\end{equation}
therefore leading to the following values of respective partial
derivatives of the Frenet frame
\begin{equation}
{\partial}_{t}\vec{t}=-{\tau}{\kappa}[1-{\kappa}{\tau}^{-2}\frac{v_{\theta}}{r}]\vec{n}
\label{20}
\end{equation}
\begin{equation}
{\partial}_{t}{\vec{n}}={\tau}{\kappa}[1-{\kappa}\vec{\tau}^{-2}\frac{v_{\theta}}{r}]\vec{t}+\frac{v_{\theta}}{r}\vec{b}
\label{21}
\end{equation}
\begin{equation}
{\partial}_{t}{\vec{b}}={\kappa}{\tau}^{-1}\frac{v_{\theta}}{r}\vec{n}
\label{22}
\end{equation}
where use has been made of the hypothesis that $\dot{\vec{b}}=0$ or
${\kappa}'(t,s)=0$, which means that the curvature only depends on
time. An important vectorial expressions is
\begin{equation}
{\partial}_{t}\vec{e_{\theta}}={\tau}_{0}sin{\theta}\vec{t}-[{\theta}_{\theta}+{\tau}_{0}]cos{\theta}\vec{n}
- [{\omega}_{\theta}+{\tau}_{0}]sin{\theta}\vec{b} \label{23}
\end{equation}
note that in the mean field dynamo case, where
$\vec{v}=\vec{v}(\vec{B})$ , equation (\ref{6}) is an eigenvalue
problem equation. Dynamo operators and eigenvalue of dynamos in
compact Riemannian manifolds have been previously investigated by
Chicone and Latushkin \cite{11}. Before presenting the field
equations for the magnetic components , we assume to simplify
matters that the ratio between them is
\begin{equation}
\frac{B_{\theta}}{B_{s}}=\frac{{\kappa}_{0}}{{\tau}_{0}} \label{24}
\end{equation}
Substitution of previous equations into equation (\ref{6}) and
splitting these equations along the components of the Frenet frame
$(\vec{t},\vec{n},\vec{b})$ yields the following three scalar
equations
\begin{equation}
{\partial}_{t}{B_{s}}+sin{\theta}{\kappa}_{0}B_{\theta}=\lambda{\alpha}B_{s}+({\partial}_{r}{\alpha})B_{\theta}
\label{25}
\end{equation}
\begin{equation}
{\partial}_{t}B_{\theta}=\lambda{\alpha}B_{\theta}+({\partial}_{r}{\alpha})B_{s}
\label{26}
\end{equation}
where  we have used the approximation
\begin{equation}
{\Omega}_{0}+{\tau}_{0}=0 \label{27}
\end{equation}
where ${\Omega}_{0}$ is the vorticity along the magnetic tube axis.
To obtain the solution one computes the dynamo factor ${\alpha}$ one
needs the equations for the dynamo flow vorticity
\begin{equation}
{\omega}_{r}=-{\partial}_{s}v_{\theta}={\kappa}_{0}{\tau}_{0}rv_{\theta}sin{\theta}
\label{28}
\end{equation}
where we have used in this equation the physical assumption of the
incompressibility of the dynamo flow. The remaining vorticity
expressions are
\begin{equation}
{\omega}_{\theta}=-{\partial}_{r}v_{s} \label{29}
\end{equation}
\begin{equation}
{\omega}_{s}={\partial}_{r}v_{\theta}+\frac{1}{r}v_{\theta}
\label{30}
\end{equation}
This allows us to obtain the following value for ${\alpha}$
\begin{equation}
{\alpha}(r)={\Omega}_{0}[{\Omega}_{0}-2r]
 \label{31}
\end{equation}
Substitution of these equations into equations (\ref{25}) and
(\ref{26}) yields the toroidal component
\begin{equation}
B_{s}=B_{0}(r)\frac{{\kappa}_{0}}{{\beta}}exp[{\beta}t] \label{32}
\end{equation}
where $\beta:={{\tau}_{0}}^{2}[\frac{{\kappa}_{0}}{{\tau}_{0}}
-\lambda]$. Thus the value for the poloidal component as
\begin{equation}
B_{\theta}=
\frac{{\tau}_{0}}{{\kappa}_{0}}B_{0}(r)\frac{{\kappa}_{0}}{{\beta}}exp[\lambda{\alpha}+
\frac{{{\tau}_{0}}^{3}}{{\kappa}_{0}}t] \label{33}
\end{equation}
where we have already used the approximation that we are very close
to the magnetic flux tube axis (r=0). By equating the time
exponentials, since the ratio between poloidal and magnetic
components does not depend on time one is able to determine
$\lambda$ as
\begin{equation}
{\lambda}=\frac{1}{2}[\frac{{{\kappa}_{0}}^{2}-{{\tau}_{0}}^{2}}{{\tau}_{0}{\kappa}_{0}}]
\label{34}
\end{equation}
Toroidal component is finally given by
\begin{equation}
B_{s}=B_{0}(r)\frac{{\kappa}_{0}}{{\beta}}exp[\frac{{\kappa}_{0}}{2{\tau}_{0}}t]
\label{35}
\end{equation}
Note that to pursue dynamo action we must have tha curvature and
torsion of the tube axis must possess the same sign or if we change
the torsion sign dynamo action must change the curvature of the axis
of the tube dynamo. To determine finally the function $B_{0}(r)$ one
uses the expression for the force-free magnetic field equation
(\ref{7}) yields
\begin{equation}
{\partial}_{r}B_{0}+[\frac{1}{r}-\frac{{\lambda}^{2}}{r}]B_{0}=0
\label{36}
\end{equation}

which yields
\begin{equation}
{B}_{0}= \frac{c_{0}}{r}\label{37}
\end{equation}
where $c_{0}$ is an integration constant. Thus the complete solution
is
\begin{equation}
B_{\theta}=
{{\tau}_{0}}\frac{c_{0}}{r}{{\beta}^{-1}}exp[\frac{{\kappa}_{0}}{2{\tau}_{0}}t]
\label{38}
\end{equation}
Note however that the  while which decays slower as we go away from
the magnetic axis of flux tube dynamo.
\section{Conclusions}
In conclusion, we have used an approximate method of first-order
torsion to find near the magnetic axis of the dynamo flux tube to
analytical (non-numerical) solutions of ${\alpha}$-dynamo equation.
Since the tube we used here is twisted, it is also non-axisymmetric,
and Cowling 1934 antidynamo theorem does not apply here. Previously
Ruzmaikin et al \cite{1} have found solutions of turbulent dynamos
where unbounded magnetic field could decay at least as $r^{-3}$
which is distinct to the bounded case of twisted magnetic flux tube
helical dynamo we have found here. The general equations of the
helical dynamo found in section II could be used to find out more
general solutions, with less degree of approximation that we used
here, for example by letting the magnetic poloidal component vary
with time and also considered the case of untwisted tubes. One
finally could say that one important consequence of the
${\alpha}$-effect in the flux tube dynamo is to change signs of
torsion and curvature simultaneously. The radial component of the
magnetic field was also dropped since we have assumed that the
cross-section of the tube is constant. These considerations can be
addressed elsewhere.
\section*{Acknowledgements}
I would like to dedicate this paper to the memory of Professor
Vladimir Tsypin, friend and teacher, a great physicist and
mathematician of plasmas, which taught me a great deal of the
applications of the Riemannian geometry to plasma physics. on the
occasion of his senventh birthday. I would like also to thank CNPq
(Brazil) and Universidade do Estado do Rio de Janeiro for financial
supports.

\end{document}